\documentclass[acmsmall,screen,nonacm]{acmart}

\usepackage{changepage}
\usepackage{listings}
\usepackage{lineno}
\usepackage{xcolor}

\usepackage{subcaption}
\usepackage[most]{tcolorbox}
\newtcolorbox{rqbox}[1]{colback=gray!15, colframe=gray!50, fonttitle=\bfseries, title={#1}, boxrule=0.5pt, arc=2pt}

\newcommand{\red}[1]{{\color{red}{#1}}}
\newcommand{\blue}[1]{{\color{blue}{#1}}}
\newcommand{\green}[1]{{\color[HTML]{009900}{#1}}}

\usepackage{xspace} 
\newcommand{\spacetime}{\textit{SpaceTime}\xspace}
\newcommand{\circled}[1]{\raisebox{.5pt}{\textcircled{\raisebox{-.9pt}{#1}}}}

\renewcommand\footnotetextcopyrightpermission[1]{}

\title{SpaceTime Programming: Live and Omniscient Exploration of Code and Execution}

\AtBeginDocument{%
  }

\newcommand{\commonaffiliation}{%
  \affiliation{%
    \institution{Université de Rennes, Inria, CNRS, IRISA}%
    \city{Rennes}%
    \country{France}%
  }%
}

\author{Jean-Baptiste Doderlein}
\orcid{0000-0002-9741-8571}
\affiliation{%
    \institution{Université de Rennes, IRISA}%
    \city{Rennes}%
    \country{France}%
  }%

\author{Djamel Eddine Khelladi}
\orcid{0000-0002-2218-650X}
\affiliation{%
    \institution{Université de Rennes, CNRS, IRISA}%
    \city{Rennes}%
    \country{France}%
  }%

\author{Mathieu Acher}
\orcid{0000-0003-1483-3858}
\commonaffiliation

\author{Benoit Combemale}
\orcid{0000-0002-7104-7848}
\commonaffiliation

\begin{document}

\title{SpaceTime Programming: Live and Omniscient Exploration of code and execution}



\begin{abstract}
Programming environments typically separate the world of static code from the dynamic execution of programs. Developers must switch between writing code and observing its execution, often with limited tools to understand the relationship between code changes and runtime behavior.
Several paradigms and approaches exist to bridge this gap, including exploratory programming for comparing code variants, live programming for immediate feedback, and omniscient debugging for exploring execution history. However, existing solutions tend to focus on specific aspects and one specific paradigm rather than providing a fully integrated environment with multiple capabilities. 
This paper introduces \spacetime Programming, a novel approach that unifies these paradigms to create a programming model for exploring both code modifications and execution flow. At the core of our approach is a trace mechanism that captures not only execution state but also the corresponding code changes, enabling developers to explore programs in both space (code variants) and time (execution flow).
As a proof of concept, we implemented a Python library supporting SpaceTime Programming and applied it in two contexts: a live omniscient debugger and a Pygame game development tool, showcased through a Flappy Bird–like game. 
We further evaluated SpaceTimePy on five real-world Python projects, finding performance overhead ranging from 35\% to 150\% on test suites.
\end{abstract}

\maketitle

\section{Introduction}
\label{sec:introduction}

Software development traditionally imposes a cognitive barrier between static code and its dynamic execution. During development, developers write code that is intended to be executed, constantly navigating between these two worlds. This duality creates a fundamental challenge: developers must mentally project how their static code will behave when executed, a cognitively demanding and error-prone process \cite{10.5555/28446.28458}.

To reduce this cognitive gap, various approaches have emerged over time. Developers frequently use simple print statements to verify code accessibility or the value of a variable at a specific moment \cite{10.1145/3180155.3180175}. More sophisticated tools, such as debuggers, allow setting breakpoints, executing the program step-by-step, or even running arbitrary code during execution \cite{Kotok1961,Stallman2011}. Some environments offer even more advanced features with omniscient debuggers that build a complete trace, allowing exploration of program execution through time \cite{bousse_omniscient_2018}. Other paradigms also attempt to reduce this barrier. Live programming offers immediate feedback
, where each code modification leads to an immediate change in execution, like using probes \cite{doderlein_liverec_2024} that continuously exhibit the state of variables in the code \cite{mcdirmid_usable_2013, niephaus_example-based_2020}. In parallel, exploratory programming focuses more on exploring different alternatives for a value or code area, allowing visualization and comparison of different execution results \cite{rein_exploratory_2018}, which is useful in various scenarios such as trade-off analysis.

These tools can be understood as supporting two fundamental dimensions of program exploration: \textbf{\emph{time}}, representing the evolution of the program state during execution, and \textbf{\emph{space}}, representing the set of possible code variants or design alternatives. However, current tools typically focus on one dimension at a time. Omniscient debuggers provide deep temporal exploration but offer limited support for modifying code, whereas live and exploratory environments emphasize rapid iteration but provide only partial access to execution history. As a consequence, these two forms of exploration are treated as complementary, yet, largely independent. We argue that bridging them (i.e., enabling exploration of both time and space simultaneously) could offer richer and more integrated forms of reasoning about program behavior.

\begin{figure}[t]
    \centering
    \includegraphics[width=0.70\textwidth]{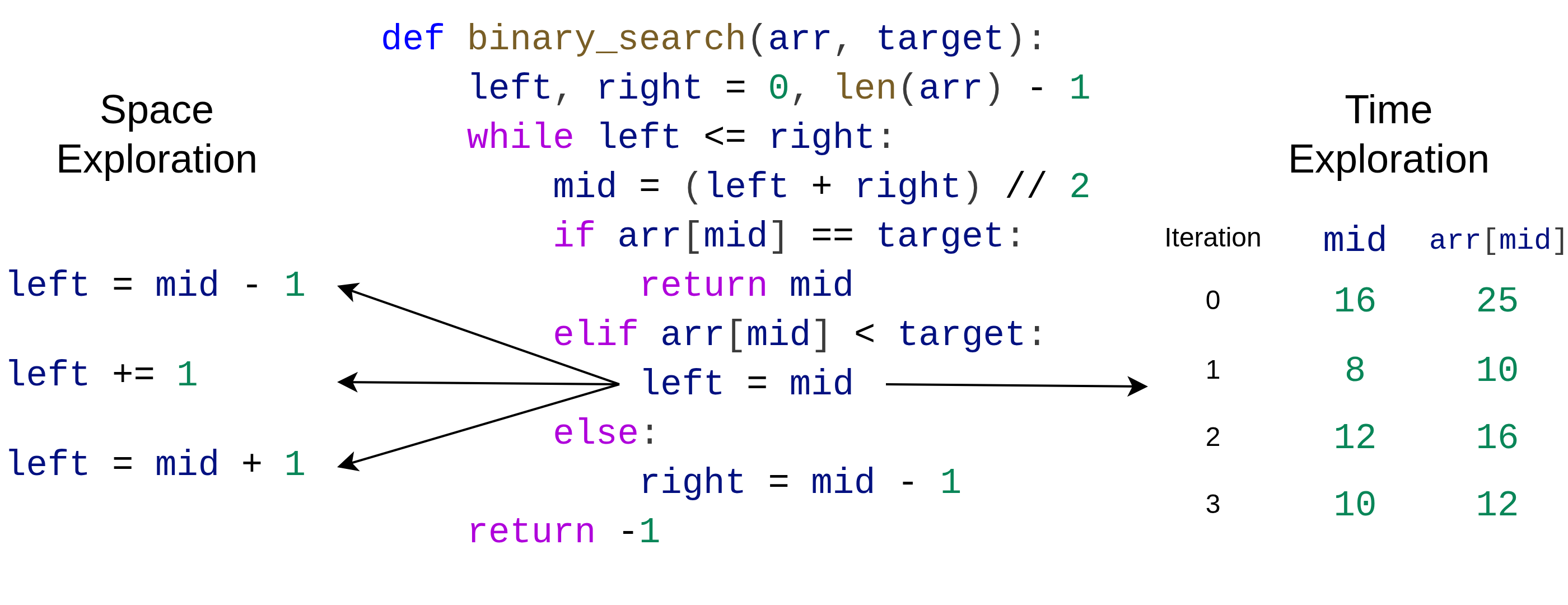}
    \caption{Example of exploration of time (states of the program) and space (variants of code) on a binary search function.}
    \Description{Diagram of a binary search development scenario showing two dimensions: time as a sequence of execution states and space as alternative code variants.}
    \label{fig:spaceandtime}
\end{figure}

To illustrate this gap, consider the development of a simple binary search function in Figure~\ref{fig:spaceandtime}. Developers may engage in two types of activities:
\begin{itemize}
\item Temporal exploration, where one analyzes execution evolution to understand how variables change over time.
\item Spatial exploration, where one modifies the code to and observes the resulting behavior.
\end{itemize}

Unfortunately, existing tools make it difficult to perform both explorations concurrently, such as comparing, over time, how different code variants affect execution. 

To address these limitations, we propose a novel unified model of \spacetime programming, grounded in a trace abstraction that captures both program state and associated code variants. This trace serves as a foundation for constructing domain-specific tools that can interactively explore execution across time and space. We demonstrate this approach through an extensible tracing system implemented in Python. Developers can define what constitutes a program state and control the level of granularity, facilitating the construction of custom visualizations and analysis tools.

We first validate the technical feasibility of our approach through two distinct prototypes built on top of this trace system. The first is a development environment for Pygame applications, supporting temporal and spatial exploration within game sessions. The second is a live omniscient debugger for Python functions, integrated with Visual Studio Code, combining execution replay with live code modification. We further evaluate the runtime performance of \spacetime and the impact of its tracing mechanism on benchmark suites and five real-world Python projects. Results show that function-level tracing introduces a moderate overhead on HumanEval dataset, while measurements on real-world Python projects range from 35\% to 150\% depending on the traced workload and captured state. In this work, we focus on demonstrating the technical viability of the \spacetime Programming model through these prototype implementations and performance evaluations. Evaluating usability at a larger scale is out of scope herein and is left for future work.

The paper is structured as follows: Section~2 reviews the background on exploratory programming, live programming, and omniscient debugging. Section~3 introduces the \spacetime Programming model, its motivating example, and its requirements. Section~4 defines the underlying trace abstraction and the operations it supports. Section~5 presents SpaceTimePy, our Python implementation of the model. Section~6 evaluates the approach through two case studies and a performance analysis. Section~7 discusses implementation trade-offs and usability considerations, Section~8 positions our work with respect to related research, and Section~9 concludes.

\section{Background}
\label{sec:background}

\begin{figure}[b] 
    \captionsetup[subfigure]{labelformat=empty,justification=centering}
    \begin{subfigure}{0.24\textwidth}
        \centering
        \includegraphics[width=\linewidth]{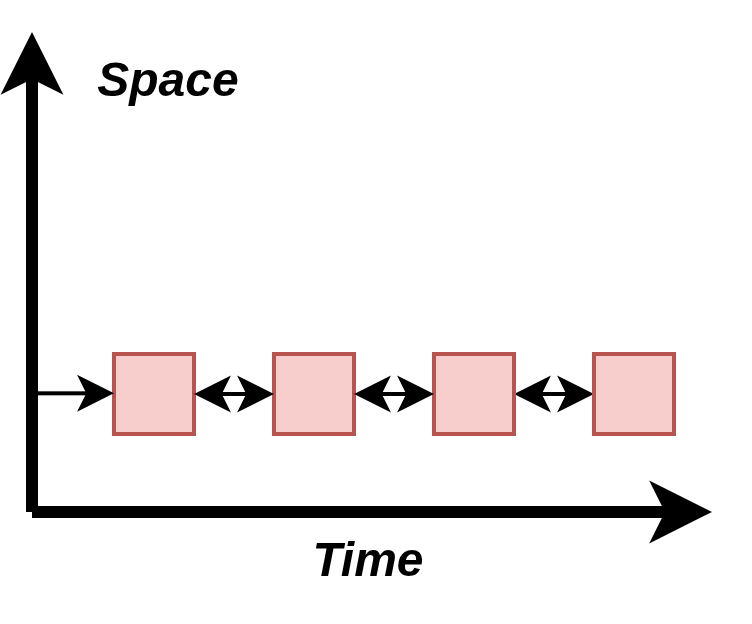}
        \caption{(a)\\Omniscient Debugging}
        \label{fig:compback_omni}
    \end{subfigure}
    \hfill
    \begin{subfigure}{0.24\textwidth}
        \centering
        \includegraphics[width=\linewidth]{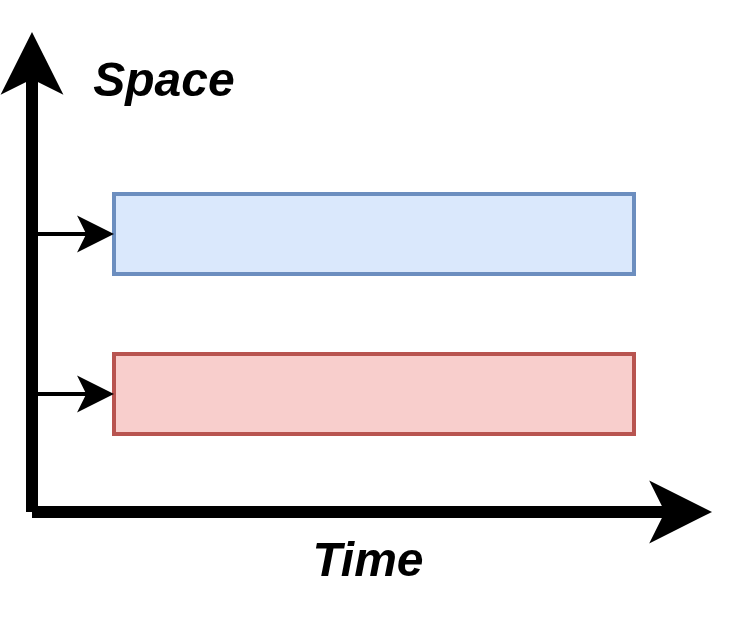}
        \caption{(b)\\Exploratory Programming}
        \label{fig:compback_explo}
    \end{subfigure}
    \hfill
    \begin{subfigure}{0.24\textwidth}
        \centering
        \includegraphics[width=\linewidth]{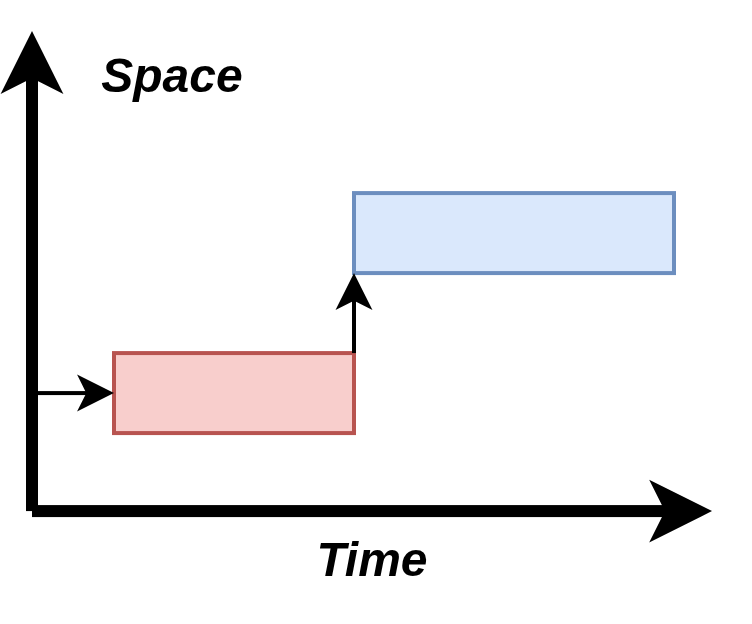}
        \caption{(c)\\Live Programming}
        \label{fig:compback_live}
    \end{subfigure}
    \hfill
    \begin{subfigure}{0.24\textwidth}
        \centering
        \includegraphics[width=\linewidth]{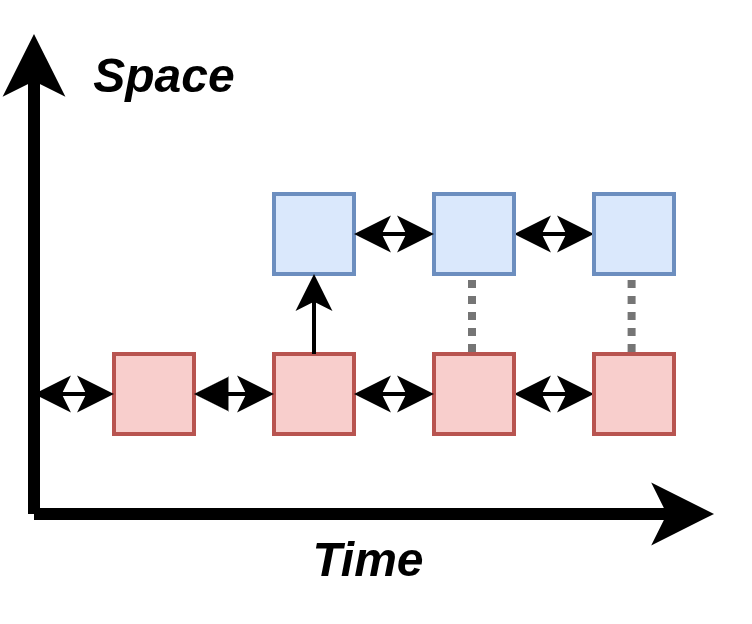}
        \caption{(d)\\SpaceTime Programming}
        \label{fig:compback_spacetime}
    \end{subfigure}

    \caption{Exploration through time and space: current approaches address either the space dimension (exploratory/live programming) or the time dimension (omniscient debugging), but not both simultaneously}
    \Description{Four-panel comparison of programming approaches: omniscient debugging focuses on time, exploratory and live programming focus on space, and SpaceTime combines both dimensions.}
    \label{fig:compback}
\end{figure}

This section presents the different approaches that allow developers to explore the space and time of a program. For each approach, we represent how they interact with space and time in Figure~\ref{fig:compback}.

\subsection{Exploratory Programming}
\label{sec:exploratory-programming}

Exploratory programming focuses on supporting developers in systematically managing and exploring multiple code variants during the development process. This paradigm recognizes that software development often involves investigating several alternative implementations, parameter configurations, or design approaches before converging on a final solution. The core challenge addressed by exploratory programming is helping developers organize, compare, and reason about these different code variants efficiently~\cite{rein_exploratory_2018}.

The fundamental principle of exploratory programming is variant management: providing developers with mechanisms to create, modify, and navigate between different variants of their code while maintaining awareness of the relationships and differences between variants~\cite{hartmann_design_2008}. Traditional development workflows force developers to manage variants manually through techniques like commenting out code sections, creating multiple files, or using version control branches—approaches that become cumbersome and error-prone as the number of variants increases.

Modern exploratory programming environments address these limitations through dedicated support for variant creation and comparison. Interactive notebook environments like Jupyter provide cell-level variant management, where developers can duplicate and modify individual code cells to explore different approaches within the same document~\cite{kery_story_2018}. Some systems extend this concept by supporting explicit variant trees and parallel development of alternative solutions~\cite{terry_variation_2004}, or providing tools to manage the complexity of iterative exploration~\cite{head_managing_2019}.

More sophisticated approaches include specialized variant management systems that track the evolution of code experiments and provide structured ways to organize exploration sessions~\cite{beckmann_probing_2025}. These systems typically maintain metadata about each variant, such as its purpose, performance characteristics, or relationship to other variants, enabling developers to make informed decisions about which approaches to pursue further~\cite{beth_kery_exploring_2017}.

Exploratory programming thus addresses the space dimension of program development by providing systematic support for managing and navigating the landscape of possible code implementations, enabling developers to efficiently explore the solution space before committing to specific approaches. This approach is represented in Figure~\ref{fig:compback_explo}, where the developer can compare multiple executions across space.

\subsection{Live Programming}
\label{sec:live-programming}

Liveness in programming environments refers to the ability to modify a program while it is running and have those changes take effect immediately~\cite{tanimoto_perspective_2013}. Live programming environments embody this concept by enabling continuous execution with real-time updates. Hence, eliminating the traditional edit–compile–run cycle and allowing developers to instantly observe how code changes affect a program’s behavior and state. The primary goal is to reduce the barrier between code and execution by providing immediate feedback on modifications, making program behavior more transparent and responsive during development~\cite{lerner_projection_2020}.

Historically, live programming emerged from dynamic environments like Lisp and Smalltalk, which pioneered features like runtime code replacement and program inspection~\cite{hancock_real-time_2003}. Modern implementations span various domains: browser developer tools enable real-time HTML/CSS modifications, JavaScript frameworks offer Hot Module Replacement (HMR) for rapid development cycles, and specialized tools like YinYang provide probe-based live programming interfaces~\cite{mcdirmid_usable_2013}.

Example-based live programming with probes was introduced by Edwards~\cite{edwards_example_2004}, who proposed integrating examples directly into the development environment. Recent implementations include language-agnostic solutions like LiveRec, which uses the Debug Adapter Protocol~\footnote{microsoft.github.io/debug-adapter-protocol} to insert probes and enable dynamic code reloading~\cite{doderlein_liverec_2024}, as well as systems described by Niephaus et al.~\cite{niephaus_example-based_2020} that use GraalVM and LSP features to build example-based live programming environments.

Live programming, like exploratory programming environments, allow developers to explore the space dimension of programming. Represented in Figure~\ref{fig:compback_live}, the developer can change the code (space) while the program is running.

\subsection{Omniscient Debugging}
\label{sec:omniscient-debugging}

Omniscient debugging, also known as time travel debugging, represents an enhancement of traditional debugging environments~\cite{lewis_debugging_2003}. The principle is to offer users the ability to navigate through the execution history of a program bidirectionally: unlike traditional debuggers that only allow forward stepping, time travel debugging enables backward navigation through execution states.

This approach is based on constructing a comprehensive trace that captures program state at various execution points, allowing developers to revisit the program's state during previous events~\cite{bousse_omniscient_2018}. The goal is to reduce developer effort by eliminating the need to restart execution to investigate earlier program states.

Omniscient debuggers have been implemented in various programming languages, each providing the ability to record and navigate program execution over time. For example, Boothe’s work on efficient algorithms for bidirectional debugging introduced an omniscient debugger for C~\cite{boothe_efficient_2000}. This debugger enables developers to move both forward and backward through program execution by leveraging event counters and checkpoint mechanisms. In the Java ecosystem, Georges et al. developed JaRec, a portable record/replay system that supports omniscient debugging for multithreaded Java applications by recording synchronization events and replaying them deterministically~\cite{georges_jarec_2004}. These implementations focus primarily on capturing the temporal evolution of program state, allowing developers to revisit and inspect previous execution points.

Bousse et al. propose a multidimensional debugger for DSLs, which allows not only time navigation but also value-history exploration.~\cite{bousse_omniscient_2018}. Their system introduces the concept of multidimensional debugging, which not only allows navigation through time but also enables exploration of value histories and relationships between different execution dimensions. By integrating these capabilities, their debugger supports richer analysis and comprehension scenarios, extending the traditional notion of omniscient debugging to encompass multiple axes of exploration.

Another related area is multi-verse debugging, which addresses program exploration by enabling the simultaneous analysis of multiple executions~\cite{torreslopez_et_al:LIPIcs.ECOOP.2019.27}. This technique is often motivated by non-deterministic program behavior, such as concurrency, where different execution paths or thread interleavings may lead to distinct outcomes. While complementary, in this paper we focus on the comparison of multiple executions that are obtained from different versions of the code and therefore fall within the scope of exploratory programming.

Time travel debugging thus proposes to explore a different axis from live and exploratory programming, i.e., the time dimension. The Figure~\ref{fig:compback_omni} illustrates how omniscient debugging allows the developer to explore the time dimension, but not the space dimension.

\section{SpaceTime Programming}
\label{sec:spacetime-programming}

While live programming, exploratory programming, and omniscient debugging each allow exploration along a single dimension, their combination enables forms of interaction that do not arise when these dimensions are considered independently. We illustrate this with a motivating example.

\subsection{Motivating example}
\label{sec:motivating-example}

Consider the development of a Flappy Bird–like game. The player controls a bird flying through pipes positioned at varying heights. Pipe positions are randomly generated, making the execution partially non-deterministic. A developer is working on two tasks: \textbf{\emph{1)}} Adding a high-score display, and \textbf{\emph{2)}} Adjusting gameplay parameters such as bird gravity to change the game difficulty.

After implementing the high-score feature, the developer observes a layout defect: when the score reaches two digits, the text overlaps another interface element. This issue manifests only after the player survives long enough to exceed a score of ten. With existing approaches, reaching this situation requires repeatedly replaying the game from its initial state. Because execution is partially non-deterministic, reproducing a similar gameplay context may be time-consuming.

If exploration in time and space are treated independently, the developer may either inspect past executions without modifying them, or modify the code but restart execution from the beginning. When these dimensions are combined, however, a different workflow becomes possible: the developer can revisit a previously recorded execution, select a state immediately preceding the problematic score threshold, modify the rendering logic, and resume execution from that state under the new code variant. In this setting, execution history becomes a persistent structure from which alternative futures can be explored. The developer no longer needs to reconstruct the context in which the modification is relevant. Instead, that context can be directly reused and transformed.

A related need arises when adjusting gameplay parameters. For example, modifying the bird’s gravity alters its trajectory and thereby the overall difficulty of the game. Assessing such a change requires observing how behavior unfolds over time. Running the program under different parameter values yields distinct execution histories, yet comparing these histories remains challenging: differences are distributed across time, influenced by non-determinism, and often difficult to recall or articulate precisely. Treating space and time jointly allows the developer to navigate multiple execution histories corresponding to different code variants, to align them at selected moments, and to compare their evolution. Such comparison may involve differences in internal state (e.g., position or velocity variables) or in rendered output. In this way, exploration shifts from examining how one program evolves to examining how alternative programs evolve differently over time.

These scenarios illustrate that integrating time and space does not only aggregate the capabilities of existing techniques. Rather, it supports development activities in which execution histories can be revisited, branched, and compared across code variants, enabling forms of reasoning that are otherwise difficult or impractical with current tools. The Figure~\ref{fig:compback_spacetime} depicts the proposed \spacetime Programming approach, which unifies both dimensions by enabling simultaneous exploration of multiple code variants across different execution steps over time.

\subsection{Requirements for SpaceTime Programming}
\label{sec:requirements}

To enable exploration of both time and space dimensions in program development, we distinguish two categories of requirements for a \spacetime Programming system. The first category covers capabilities inherited from exploratory programming, live programming and omniscient debugging.

\begin{itemize}
    \item \textbf{Temporal Exploration}: Explore previous execution states to understand program behavior by navigating execution history and inspecting program state at any point in time.
    \item \textbf{Code Variant Exploration}: Define different code variants and compare their execution.
    \item \textbf{Live Interaction}: Apply code modifications to running examples with immediate feedback.
\end{itemize}

Their combination introduces two additional capabilities:
\begin{itemize}
    \item \textbf{Temporal Comparison of Variants}: Compare multiple variants over time by observing their execution evolution simultaneously and comparing their states at corresponding steps.
    \item \textbf{Live Execution Branch Exploration}: Use previous execution states as entry points and live program within a selected time frame with captured external interactions.
\end{itemize}

To address these requirements for a system allowing navigation through time and space, we propose to use an extensible trace structure. This trace system is a data structure that captures both runtime execution data, associated code and external interaction.

\subsection{SpaceTime Development Workflow}
\label{sec:spacetime-workflow}

\begin{figure*}[!ht]
    \centering
    \includegraphics[width=0.9\textwidth]{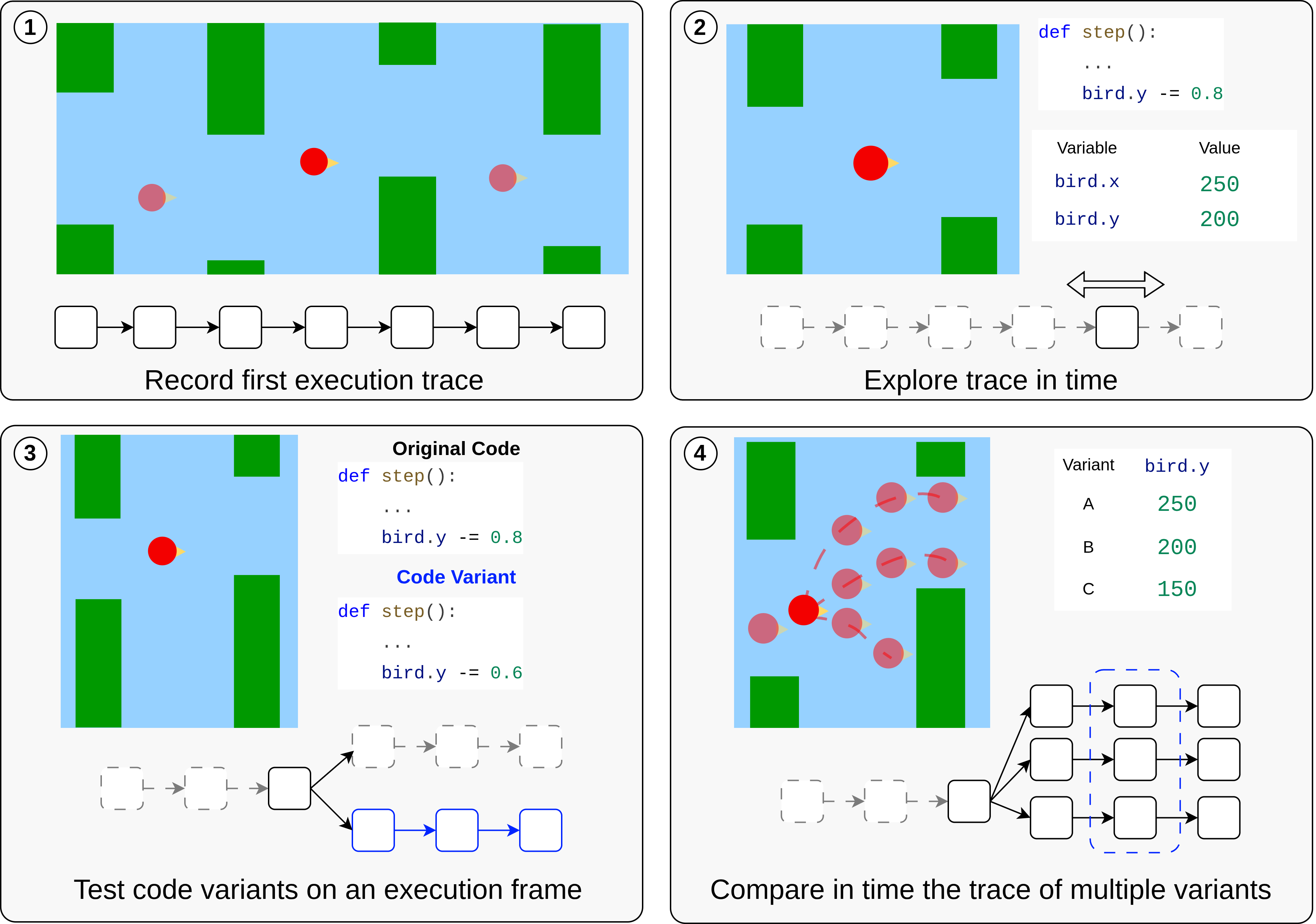}
    \caption{The \spacetime development workflow in four steps. \circled{1}~The developer runs the program and records an execution trace. \circled{2}~The developer selects an execution frame (a sub-portion of the trace) and inspects states to understand program behavior over time. \circled{3}~The developer modifies the code and replays the selected execution frame with the new code variant $C'$, while mocking external events from the original trace. \circled{4}~The two execution branches are aligned in time and compared state by state, allowing the developer to observe how the code change affects behavior within the same execution context.}
    \Description{Workflow diagram with four numbered steps: record a trace, inspect a selected frame, replay with modified code and mocked events, then compare aligned original and replayed executions.}
    \label{fig:spacetime_workflow}
\end{figure*}

We now describe the \spacetime workflow shown in Figure~\ref{fig:spacetime_workflow}, using Flappy Bird as a running example. The workflow has four iterative steps.

First, the developer runs the program with tracing enabled (\circled{1} in Figure~\ref{fig:spacetime_workflow}). The system captures a sequence of states $s_1, s_2, \dots, s_n$, each containing variables, external events, and the associated code version. In our Flappy Bird example, this corresponds to recording a complete game session where each frame produces one state in the trace.

The developer then navigates the recorded trace to understand program behavior over time (\circled{2}). The developer can select an execution frame: a sub-portion of the trace corresponding to a specific moment of interest. For instance, the developer may focus on the frames where the bird approaches a narrow pipe passage. Within this frame, each state can be inspected to observe variable values, the executed code, and any external events that occurred. This step corresponds to the temporal exploration requirement.

With a relevant execution context identified, the developer modifies the source code and triggers a replay on the selected execution frame (\circled{3}). For example, changing gravity and replaying creates a new execution branch, using the new code variant $C'$ while mocking the external events captured in the original trace. This produces a new sequence of states that reflects the modified code under the same execution context. The developer does not need to replay the entire game from the beginning or manually recreate the conditions that lead to the moment of interest.

Finally, the developer compares the original and replayed branches within the selected frame (\circled{4}), inspecting aligned states (or visual overlays) to see how the gravity change alters behavior. This step addresses temporal comparison of variants.

This workflow is iterative: after comparison, the developer may select a different execution frame, try another code modification, or refine the current one, progressively converging toward the desired behavior. The trace serves as a persistent, reusable artifact from which multiple exploration branches can be derived without re-executing the full program.

\subsection{Challenges}
\label{sec:spacetime-challenges}

Realizing the requirements of Section~\ref{sec:requirements} in a single system raises a set of technical challenges. We summarize them here as they motivate the trace-based design presented in the following section.
The challenges first come from the tools and paradigms that inspire \spacetime :

\begin{itemize}
    \item \textbf{Trace cost vs. fidelity} : capturing enough information to inspect past states, while controlling runtime overhead, trace size, and serialization costs.
    \item \textbf{Hot code replacement} : loading modified code and continue execution while preserving a coherent context and maintaining short feedback cycles.
    \item \textbf{Variant management and comparison} : managing multiple code variants and their outputs.
\end{itemize}

However, as for the requirements, combining these capabilities introduces additional challenges that arise specifically from their interaction:

\begin{itemize}
    \item \textbf{State reinstantiation under code substitution}: reinstating a recorded state as a replay entry point while substituting code at the chosen granularity.
    \item \textbf{External interaction handling}: capturing and reproducing relevant environment-dependent interactions so that variants can be evaluated under comparable conditions.
    \item \textbf{Cross-variant alignment}: aligning executions across time to compare corresponding steps.
\end{itemize}

To address these requirements and challenges, we rely on an extensible trace structure that captures runtime execution data, associated code versions, and recorded external interactions.

\section{Design}
\label{sec:design}
\subsection{Trace Definition}
\label{sec:trace-definition}

Our approach relies on the construction of an execution trace. A trace is composed of states, each representing a snapshot of the program at a specific instant during execution. Each state is defined by three elements:

\begin{enumerate}
    \item \textbf{Variables} ($V$): all data collected at that instant, including local variables (stack) and relevant global variables (heap).
    \item \textbf{Events} ($E$): the external events observed between this instant and the next captured state. An external event is any observable interaction between the program and its environment whose outcome is not deterministically reproducible from the source code alone, such as user input, random number generation or network responses.
    \item \textbf{Code} ($C$): the version of the source code being executed, along with its location at the chosen granularity.
\end{enumerate}

We denote a state as $s_t(V, E, C)$, where $t$ is its order of appearance in the execution. An execution trace is a temporally ordered sequence of such states.

States can correspond to different levels of granularity depending on the needs of the tool or the developer. A system supporting SpaceTime Programming must allow this granularity to be configured. For example, a coarse granularity might capture one state per function call (recording call arguments and return value), while a finer granularity might capture one state per source-level statement, enabling step-by-step inspection of how variables evolve within a function. Different granularity levels can coexist within the same trace: the system may record coarse-grained states for the overall execution while simultaneously recording finer-grained states for selected regions of interest. The trace is then naturally structured as a hierarchy, where higher-level states (function calls grouped into a session) can contain sequences of more detailed states (per-statement snapshots).

Beyond granularity, the developer must be able to specify which parts of the code are subject to monitoring. Only the designated regions incur the cost of instrumentation; the rest of the program executes without overhead. By default, the system captures all variables accessible at the chosen granularity, but the developer should be able to refine this by including or excluding specific variables, for instance to reduce trace size or to avoid capturing objects that are not serializable.

To populate the event component $E$, the developer designates specific callables whose invocations within monitored code should be recorded. For each such tracked callable, the system captures its arguments and return value whenever it is invoked during the execution of monitored code. Any callable whose behavior depends on external factors (e.g., input retrieval, random number generation) can be tracked, and its recorded results can later be reproduced during replay.

Finally, the system supports hooks: user-defined functions that execute at specific points during monitoring (e.g., at function entry or return) to compute and record additional metadata. Hooks allow the trace to include domain-specific information that is only available at runtime and cannot be reconstructed from variables alone. For instance, a hook at function return could capture a screenshot of the current display in a graphical application or compute a derived metric from the return value.

\subsection{Trace Operations}
\label{sec:trace-operations}

\begin{figure}[h]
    \centering
    \includegraphics[width=0.80\textwidth]{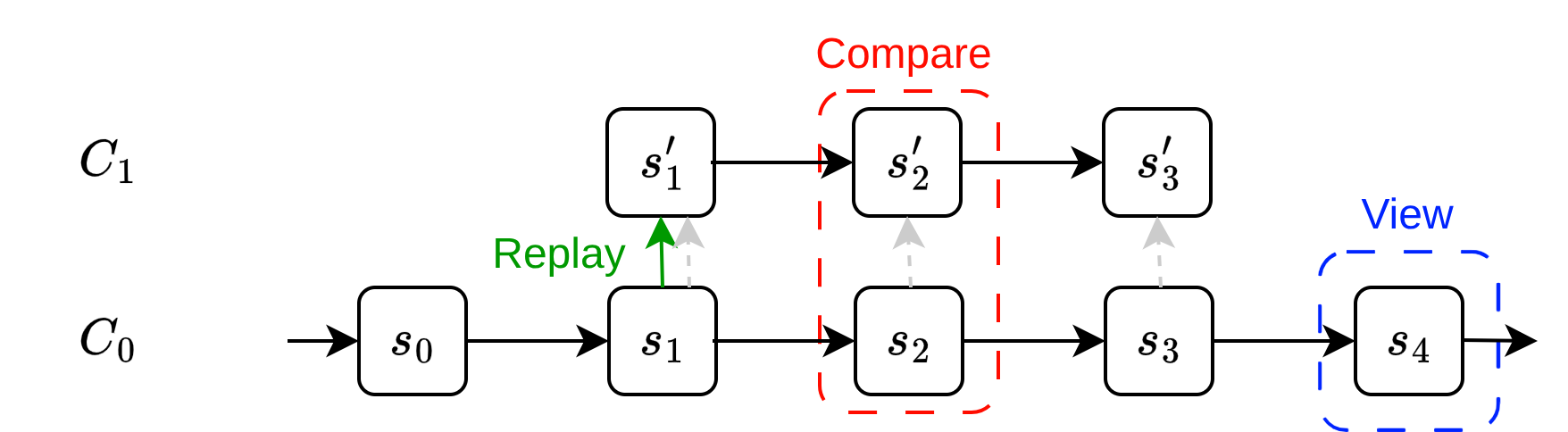}
    \caption{Illustration of the three operations defined on traces. The blue box represents the inspection of a state (\blue{$view(s_4)$}), the red box shows a comparison between different code variants (\red{$compare(s_2, s'_2)$}), and the green arrow illustrates a replay starting from a given state with replication of selected events (\green{$replay(\{s_1, s_2, s_3\},\{E_1, E_2, E_3\}, C_1)$}).}
    \Description{Trace operations diagram highlighting state inspection, cross-variant state comparison, and replay from selected states with reproduced external events.}
    \label{fig:trace_def}
\end{figure}

From a set of execution traces, we define two families of operations: inspection and replay (Figure~\ref{fig:trace_def}).

\subsubsection{Inspection}

Inspection operations allow observation of the trace. The operation $view(s_t) \rightarrow V \,|\, E \,|\, C$ provides access to the variables, events or code of a given state. The operator $compare(s_t, s'_t)$ compares two states corresponding to the same execution step but originating from different code variants. This comparison applies to any component of the state (variables, events, code) and may also extend to domain-specific metadata captured via hooks, such as rendered visual output. These two operation are represented with the blue and red box in Figure~\ref{fig:trace_def}

\subsubsection{Replay}

Replay operations launch a new execution from one or more existing trace states while controlling the context:
$$replay(\{ s_t, \dots, s_{t+i} \}, \{ E_t, \dots, E_{t+i} \}, V_t, C')$$
This operation replays a segment of an execution by reproducing selected external events $\{E_t, \dots, E_{t+i}\}$, starting from the variable state $V_t$, but using a new version of the code $C'$. The developer has four degrees of freedom when configuring a replay.

First, the developer chooses which portion of the trace to replay. Rather than replaying an entire session, a contiguous subsequence of states can be selected, defining an execution window. This limits both the computation cost and the amount of information the developer must process, which is particularly useful for focusing on a specific phase of a long-running execution.

Second, the developer can selectively choose which variables to restore from the trace and which to set manually or leave unchanged. This allows for controlled experimentation, such as restoring the program state except for a single parameter that the developer wants to modify.

Third, the developer specifies how external events are handled during replay. Tracked callables recorded during the original execution can be replaced by mocks that reproduce their recorded return values in the same order. If the replayed execution invokes a tracked callable more times than originally recorded, the system falls back to the actual callable for the additional calls. The developer can also choose to leave some tracked callables unmocked, allowing new external input during replay. This selective mocking enables deterministic replay of executions that originally depended on external factors while still allowing the developer to interact where needed.

Fourth, the developer chooses which version of the code to use. By default, replay uses the current version, but any previous version from the trace or any other variant can be supplied. Combined with event mocking and selective state migration, code substitution enables comparison of how different code variants behave under identical conditions.

A replay operation is displayed in Figure~\ref{fig:trace_def} with a green arrow, where a subset of 3 state($s_1, s_2, s_3$) are replayed with external event ($E_1, E_2, E_3$), represented as gray arrow, with a new code variant $C_1$.

These operations directly address the requirements from Section~\ref{sec:requirements}. Temporal exploration is supported by $view$ and navigation through the ordered sequence of states. Code variant exploration relies on the explicit recording of $C$ in each state and the ability to maintain traces with different code versions. Live interaction is realized through $replay$ with the current code. Temporal comparison of variants is enabled by $compare$ across aligned traces. Live execution branch exploration combines partial replay from a selected state with event mocking and code substitution.

\subsection{Required Programming Language Features}
\label{sec:required-pl-ft}
We restrict our scope to languages and execution models whose behavior can be captured as a trace.
Implementing the trace model of Section~\ref{sec:trace-definition} and the operations of Section~\ref{sec:trace-operations} requires specific technical capabilities from both the programming language and the execution environment. The targeted language and runtime must provide the following capabilities:
\begin{itemize}
    \item Observe variable values at the chosen granularity (to populate $V$ and support $view$ and replay entry points).
    \item Identify the code being executed at the same granularity (to record $C$ and enable replay with code substitution).
    \item Serialize captured values so they can be persisted in the trace.
    \item Record external interactions to later reproduce outcomes (to populate $E$ and support event mocking during replay).
    \item Load or update code dynamically inside the running environment (to replay with $C'$).
    \item Re-instantiate recorded values and resume execution from a given granularity (to branch from past states and run partial replays).
\end{itemize}

Capturing $V$ and $C$ at the chosen granularity can be achieved either by transforming code (e.g., inserting probes) or by relying on runtime instrumentation mechanisms such as debugger interfaces. Code identity and location may be obtained from compilation artifacts (e.g., debug information such as DWARF for compiled languages) or through runtime introspection in dynamic languages.

Persisting traces requires serialization support. While many ecosystems provide standard mechanisms, some runtime objects are not serializable without custom adapters (e.g., closures or resource handles), which constrains what can be stored in $V$ and what must instead be represented indirectly.

Replay requires reconstructing a past execution context: injecting recorded variable values, controlling external interactions, and running with an updated code variant $C'$. 
External events can be reproduced by recording tracked callables and replacing them with mocks that return recorded values. Replaying with code substitution further requires dynamic loading or hot-swapping support. In dynamic languages this is typically supported through late binding, whereas other environments expose more constrained hot-swap mechanisms, often at function level. The available granularity conditions both feasibility and performance.

\section{SpaceTimePy: An implementation for Python}
\label{sec:spacetimepy-implementation}

This section presents our Python implementation of the SpaceTime trace system: monitoring, trace storage, and replay with code variants.
The source code is available online.

\subsection{Monitoring Execution}
\label{sec:monitoring-execution}

\begin{figure}[h]
    \centering
    
    \begin{minipage}{0.85\textwidth}
        \centering
        \begin{lstlisting}
def move_player(x:int, y:int, player:Player) -> bool:
    player.x = x
    player.y = y
    if y < 0 or y > 600:
        return False
    if x < 0 or x > 800:
        return False
    return True
        \end{lstlisting}
    \end{minipage}
    
    \vspace{0.2em}
    
    \begin{minipage}{0.49\textwidth}
        \centering
        \begin{lstlisting}
@spacetimepy.function
def move_player(...):
    ...
        \end{lstlisting}
    \end{minipage}
    \hfill
    \begin{minipage}{0.48\textwidth}
        \centering
        \begin{lstlisting}
@spacetimepy.line
def move_player(...):
    ...
        \end{lstlisting}
    \end{minipage}
    
    \vspace{0.2em}
    
    \begin{minipage}{0.48\textwidth}
        \centering
        \includegraphics[width=0.9\linewidth]{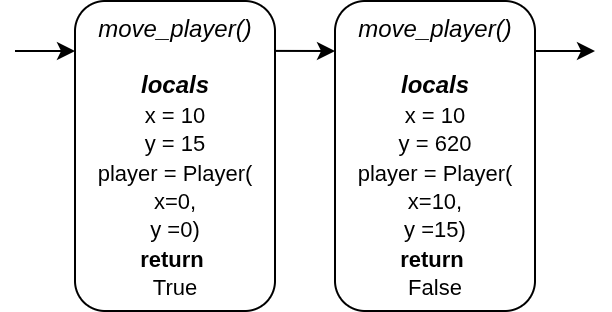}
    \end{minipage}
    \hfill
    \begin{minipage}{0.48\textwidth}
        \centering
        \includegraphics[width=0.8\linewidth]{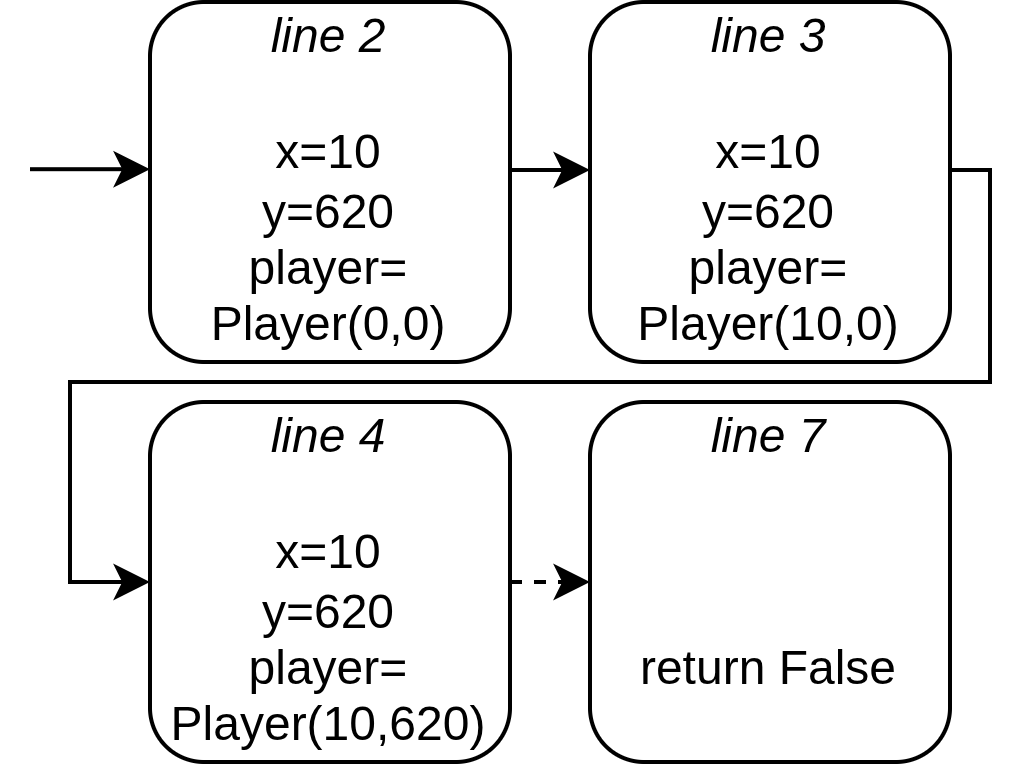}
    \end{minipage}
    
    \caption{SpaceTimePy instrumentation example for \texttt{move\_player}: function-level monitoring (left) records one state per call, while line-level monitoring (right) records one state per executed line.}
    \Description{Instrumentation example combining code snippets and two diagrams to contrast function-level monitoring with line-level monitoring and their resulting trace states.}
    \label{figure:monitoring_example}
\end{figure}

The trace system is built around a module that allows defining the specification for trace recording. An example of this specification is given in Figure~\ref{figure:monitoring_example}: \texttt{@spacetimepy} decorators are used to define the functions to track and the granularity of the trace. The \texttt{@spacetimepy.function} decorator tracks the function call and return, and the \texttt{@spacetimepy.line} decorator tracks the execution of each line of code. When executed, the tagged function will generate states depending on the chosen granularity. As shown in Figure~\ref{figure:monitoring_example}, function granularity will generate one state per function call that contains the call context and return value. For line granularity, a new state will be generated for every line in the function body (plus the calling context and return value).

\begin{figure}[ht]
    \centering
    
    \begin{minipage}{0.58\textwidth}
        \centering
        \begin{adjustwidth}{0.5cm}{0cm}
        \begin{lstlisting}
def get_events():
    # Context dependent value
    return ... 

@spacetimepy.line(
  return_hooks=[
    lambda return_value: metric(return_value)
  ],
  track=[get_events])
def main_loop():
    for event in get_events():
        process(event)
    ...\end{lstlisting}
        \end{adjustwidth}
    \end{minipage}
    \begin{minipage}{0.40\textwidth}
        \centering
        \includegraphics[width=0.8\linewidth]{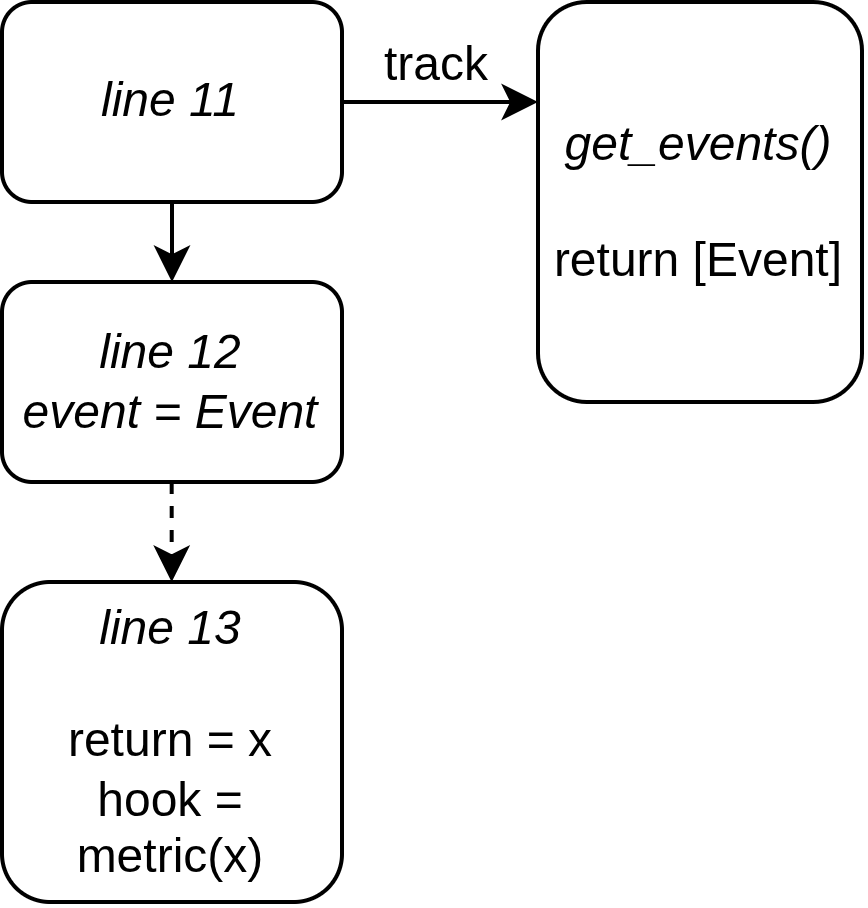}
    \end{minipage}
    \caption{Example of a monitored function with return hooks and tracked calls.}
    \Description{Code example with accompanying diagram showing a monitored main loop, a tracked external function call, and a return hook that records derived metrics in the trace.}
    \label{figure:option_example}
\end{figure}

To retrieve values during execution, we use the \texttt{sys.monitoring} module, which was added to the standard library starting with Python 3.12. This module allows defining functions to be executed during certain events occurring in the Python VM. The events of interest are \texttt{PY\_CALL}, \texttt{PY\_RETURN}, and \texttt{LINES}. \texttt{PY\_CALL} and \texttt{PY\_RETURN} correspond to function call and function return events. \texttt{LINES} captures the event of a line of code executed in the VM, which matches the granularity of a step in a debugger. An even finer granularity allows capturing bytecode instructions, but this is less useful in our case because we cannot directly link it to source code. The events are defined locally for the code and are only triggered when previously specified functions are invoked. This reduces performance impacts on the rest of the non-instrumented code.

For earlier versions of Python, similar capabilities can be achieved using the \texttt{sys.settrace} function, which has traditionally been used in debuggers. \texttt{sys.settrace} allows setting a trace function that is invoked for various events such as function calls, returns, and line executions, providing a way to monitor program execution, albeit with a different API and generally higher overhead compared to \texttt{sys.monitoring}.

These events trigger a callback that pauses code execution at the moment of the event. For \texttt{PY\_START}, the event is triggered when the function's stack frame is present on the stack, before the execution of the first line of code. For \texttt{PY\_RETURN}, the event is triggered when the function is removed from the stack. The event also provides access to the function's return value.
Finally, \texttt{LINES} is triggered before the execution of a line of code.

On these callbacks, we save the state of the program composed of local variables, global variables, domain-specific metadata, and the associated code. For function granularity, locals correspond exactly to the call arguments. To limit the number of globals recorded and only consider those useful for the function execution, we perform bytecode analysis of the function. We recursively search in the call tree for lines containing the \texttt{LOAD\_GLOBAL} bytecode. To capture code changes, we record the code associated with the function using the \texttt{inspect} module, the reflection tools present in Python.

The decorator also accepts additional parameters that provide more control over how a function is monitored. The developer can specify variables to include/exclude from recording, or add hooks that execute at the start or end of the function to capture extra metadata. There is also a parameter that allows specifying a list of functions to "track" in relation to the monitored function. When this is used, the monitoring system will record not only the main function, but also any calls to the specified tracked functions that occur within the context of the monitored function's execution. For each such call, the system captures the call context and return value, but only as they happen during the execution of the decorated function. This feature is used to capture external events, as described in Section~\ref{sec:design}.
The Figure~\ref{figure:option_example} shows an example of a function that is monitored with additional options. The \texttt{return\_hooks} parameter is used to define a function that will be executed at the end of the function call. The \texttt{track} parameter is used to define a list of functions to track. In this example, the function \texttt{get\_events} is tracked, and the return value of the function is passed to the \texttt{metric} function and added to the trace. The hook's computed value will be attached to the corresponding state, and every external event is recorded as a state that is linked to the trace.

\subsection{Trace Memory Representation}
\label{sec:storing-trace}

 \begin{figure}[h]
     \centering
     \includegraphics[width=0.8\textwidth]{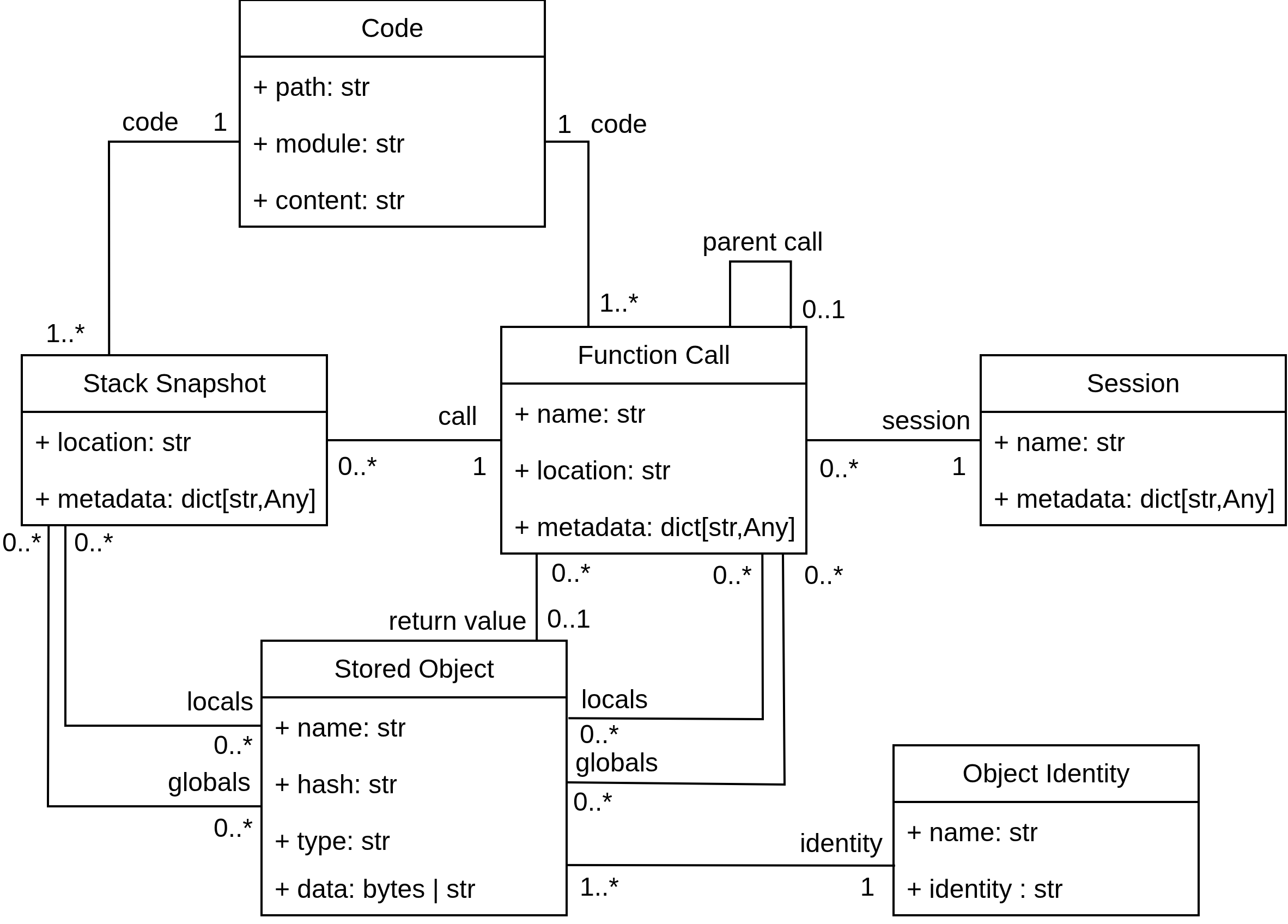}
     \caption{SpaceTimePy Trace Data Model.}
     \Description{Entity-relationship style data model for SpaceTimePy trace storage, including sessions, monitored calls, snapshots, code versions, and serialized objects.}
         \label{fig:trace_model}
 \end{figure}

To store the trace, we use the trace model shown in Figure~\ref{fig:trace_model}, which is persisted in a relational database.

This data model implements the trace definition introduced in Section~\ref{sec:trace-definition}: it represents the temporal organization of executions together with the recorded code, variable values, and associated runtime information.

The model represents the two granularities described above: function call and line execution (referred to as stack snapshots in our implementation), with an additional level, called a session, that encapsulates a time-ordered sequence of function calls.
The function model contains references to the code of the function and to the values of local variables, global variables, and return values. It also maintains a reference to a parent caller when the recording of this function occurs within the context of another monitored function (tracked function). When the granularity is set to line level, the function model contains snapshots that, like the function model itself, capture the values of local and global variables along with the associated code.

The system includes two additional models: StoredObject and Code. These represent the value of a variable and the code at a specific point in execution, respectively. These models are separated to address the redundancy issue inherent in trace collection.

An important performance and memory usage consideration is the redundancy of this data, which becomes particularly important when recording at the line level, where global variables are captured at each line of code execution despite rarely changing. To address this, we perform deduplication of recorded objects.

We distinguish between two types of equality: identity equality, which corresponds to a Python object having the same id, and content equality, which compares the actual content of objects. Since an object's content can change over time while maintaining the same identity, we also consider content equality between two objects. These two equality concepts allow us to reason about the evolution of objects over time, enabling us to track an object's history and observe the evolution of its versions.
Each object is stored in the database and assigned a unique identifier. This identifier is associated with a specific version of an object rather than the object itself. When the same object changes during execution, it receives a different identifier, allowing us to accurately capture the program state at any specific moment.
During capture, each object is compared to those already present in the database. If an equivalent object already exists, it is not stored again, and the module returns the identifier of the existing object. This mechanism both reduces the trace storage space and enables reasoning about objects over time. A similar mechanism based on textual equality is implemented to limit the storage of duplicate code.

For object serialization, we use Python's standard \texttt{pickle} module. We make two exceptions: primitive values (integers, floats, strings) are stored directly as text, and collections (lists and dictionaries) are serialized using JSON format. The values contained within these collections are serialized following the same object serialization rules.

Some objects are not serializable by default or are not properly handled by \texttt{pickle}. For these cases, we provide two options: ignore these values if they are not necessary for the trace, or provide custom serialization methods associated with the specific object types.

In our testing and evaluation, we used SQLite as the database engine. To enhance performance, we provide an option that allows the use of an in-memory database during capture, with the trace being persisted to disk upon program exit.

\subsection{Get data replay trace}
\label{sec:replay-from-trace}

\begin{figure}[t]
        \centering
        \begin{adjustwidth}{0.5cm}{0cm}
        \begin{lstlisting}
# Get data from trace
session = db.get_session()[0]
call_list = db.get_call("main_loop", session=session)
local, global, code = db.get_calling_context(call_list[0])
trace = db.get_trace(call_list[0])
# Replay from trace
spacetimepy.replay_session(session)
spacetimepy.replay_function(call_list[0])\end{lstlisting}
        \end{adjustwidth}
    \caption{Example usage of SpaceTimePy for querying and replaying traces.}
    \Description{Python code listing demonstrating trace database access, retrieval of call context and trace data, and replay of a session or a specific function call.}
    \label{figure:trace_usage_example}
\end{figure}

The SpaceTimePy module provides helper functions for accessing the content of a trace. Figure~\ref{figure:trace_usage_example} illustrates some of its capabilities. Developers can query the recorded trace database by function name, retrieve execution state or source code, and even obtain the complete trace of an execution.

The SpaceTimePy module provides utilities that enable tools to replay executions based on captured traces, as described in Section~\ref{sec:spacetime-programming}. Previous executions of functions can serve as a basis for future replays. The library exposes a method \lstinline!spacetimepy.replay_function(function_id)!, which extracts local and global variables from a previous call, uses the local variables as arguments, and restores the global variables in the environment in order to launch the function in the same context as originally captured. 

The library also exposes a method to replay an entire execution, represented as a sequence of function calls from a captured session in line 12 (\lstinline!spacetimepy.replay_session(session_id, steps)!). This function first loads the global context of the initial function, and then replays the functions recorded in the session in chronological order. The \lstinline!steps! parameter allows users to restrict the replay to a subset of the session, thus enabling partial replays of selected segments.

By default, each replay is executed using the current code version of the functions being replayed, while the global context is restored by initializing all global variables from the trace. Both methods allow the developer to substitute the executed code (for example, by using a previous code version), and to refine the reinstantiation of the context. It is possible to selectively initialize only some variables from the trace while manually setting others. 

Finally, the replay mechanism provides support for mocking functions that were marked as tracked during the original execution. When such a function is encountered during replay, it is replaced by a mock that reproduces the results recorded in the trace, returning the same values in the same order as during the original execution. If the function is invoked more times than recorded, the system falls back to executing the actual function for the additional calls. This mocking functionality allows tool developers to handle external effects separately from core program logic. For example, in a game development scenario, user input events or random number generation can be mocked to ensure consistent replay behavior, while developers focus on testing changes to game logic without being affected by the variability introduced by external factors.

\section{Evaluation}
\label{sec:evaluation}

In this section, we evaluate SpaceTimePy along three dimensions: feasibility, applicability, and performance. More precisely, we ask whether the implementation supports exploration across time and code variants, whether it can be adapted to distinct domains with manageable tool-development effort, and what runtime and storage overhead its tracing mechanism introduces.

This evaluation is guided by three research questions:
\begin{itemize}
    \item \emph{(feasibility)} \textbf{RQ1}: Does SpaceTimePy make spacetime programming feasible by supporting exploration of program execution across both time and code variants?
    \item \emph{(applicability)} \textbf{RQ2}: Can SpaceTimePy support different domains and use cases, and what tool-development effort does this require?
    \item \emph{(performance)} \textbf{RQ3}: What runtime and storage overhead does tracing introduce, and how does it affect the practical applicability of SpaceTimePy on real-world applications?
\end{itemize}

RQ1 and RQ2 are both answered in Sections \ref{sec:flappy-bird} and \ref{sec:omniscient-live-debugger}. They are mainly addressed with two tools built on SpaceTimePy, a Pygame exploration interface and a VSCode live omniscient debugger, to assess capabilities and adaptation effort. We also demonstrate the feasibility and applicability on a diversity of Python functions (coming from HumanEval) and on Github projects with significant codebases. Finally, RQ3 evaluates performance aspects and tracing overhead on HumanEval and five real-world Python projects in Section \ref{sec:performance}.

\subsection{Pygame Development, Flappy Bird game example}
\label{sec:flappy-bird}

In this case study, we apply SpaceTimePy to a Flappy Bird-like Pygame game to show capture and replay under external effects such as user input and randomness. 

\subsubsection{Pygame Trace Specification}
\label{sec:pygame-trace-additions}

\begin{figure}[!ht]
    \begin{lstlisting}
def save_screen(*args,**kwargs):
    buffer = io.BytesIO()
    pygame.image.save(pygame.display.get_surface(), buffer, "PNG")
    return {"image": base64.encodebytes(buffer.getvalue()).decode('utf-8')}

@spacetimepy.function(
        return_hooks=[save_screen],
        track=[pygame.event.get, random.randint])
def display_game():
    ... \end{lstlisting}
    \caption{Pygame Code Monitoring Configuration}
    \Description{Python code listing showing the monitored Pygame display loop with a return hook and tracked calls to input and randomness functions.}
    \label{listing:pygame_monitoring}
\end{figure}

We model a Pygame game as a main display function that renders each frame and is executed continuously throughout the game session. A single execution of this function corresponds to one frame and constitutes a trace step. We attach monitoring to this function using the \texttt{@spacetimepy.function} decorator, and define a complete game session as a monitoring session. Listing~\ref{listing:pygame_monitoring} shows the code structure and decorators used for this instrumentation.

Beyond capturing local and global variables, we augment the trace with domain-specific information to facilitate game execution analysis and enable replay with code modifications. First, we use a return hook to capture a screenshot of the game screen at the end of each frame, storing it in PNG format to optimize database size. Direct storage of the surface bitmap results in prohibitive storage requirements and performance degradation.

Second, we track two external functions that introduce non-determinism : \texttt{pygame.event.get}, which retrieves and clears the event buffer containing user inputs such as keyboard presses, and \texttt{random.randint}, which controls pipe placement generation. Together, these tracked functions capture the external effects that influence game behavior and are necessary for faithful replay.

Finally, since the Pygame library utilizes C APIs internally, we implement custom picklers to properly serialize Event objects and other Pygame-specific data structures. We also configure the system to ignore non-serializable objects such as the screen surface to prevent serialization errors.

\subsubsection{Interactive Exploration Tool}
\label{sec:interactive-exploration-tool}

\begin{figure}[h]
    \centering
    \includegraphics[width=0.75\textwidth]{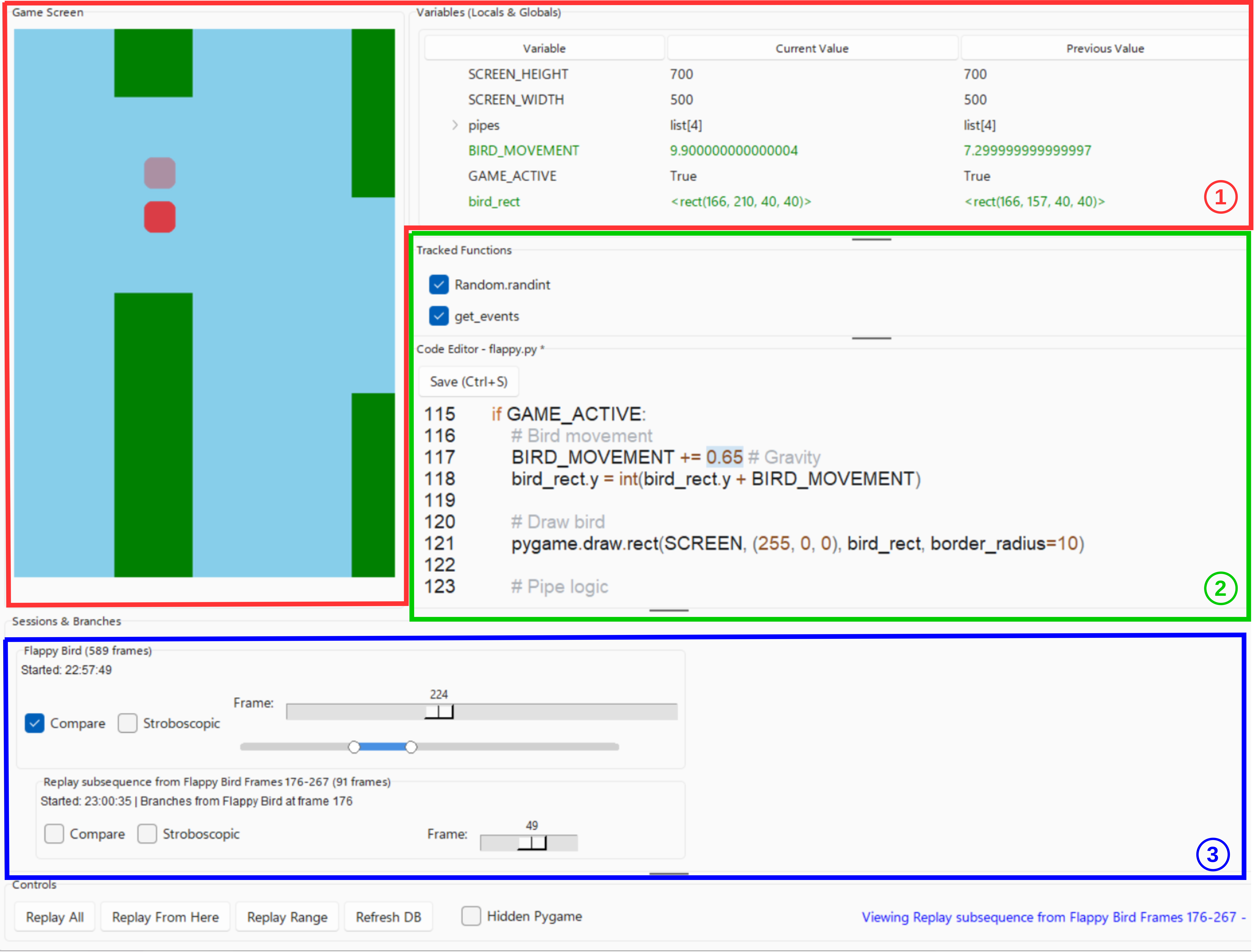}
    \caption{Interactive Flappy Bird exploration tool interface showing three main areas: (1) red zone for execution and state inspection view, (2) green zone for state migration configuration, and (3) blue zone for timeline sliders and replay controls. The interface displays one game session of 400 frames and one replay from frame 223 with changed gravity. The two executions are compared, with the semi-transparent overlay showing the original game session.}
    \Description{Screenshot of the Flappy Bird exploration interface with a game view, state migration options, timeline controls, and side-by-side comparison between original and replayed executions.}
    \label{fig:flappy_interface}
\end{figure}

We have developed an interactive tool for exploring Pygame games using our trace system. Figure~\ref{fig:flappy_interface} shows a screenshot of the visualization and replay interface.

The workflow begins with the user recording an initial game session with monitoring activated. This recorded session serves as the reference example for subsequent exploration and analysis.

The tool interface provides several key features for spacetime exploration organized into three distinct zones. The red zone (1) provides execution and state inspection capabilities, displaying the game screen and variable values at the selected time point, enabling developers to understand the game state evolution. The blue zone (3) contains timeline controls including a slider that allows users to navigate through the game's execution timeline, instantly displaying the corresponding screen capture and variable states. Additionally, three replay buttons provide different behaviour: the first one replays the entire game sequence from the beginning, the second initiates replay from the currently selected time point and the third one allows replay on a specific execution frame.

The green zone (2) offers fine-grained control  for configuration or state migration. Users can choose which global variables should be transferred from the original execution, allowing selective preservation of game state. They can also specify which tracked functions should be mocked during replay, enabling control over external effects.

When replaying without code changes and with both event and random function mocking enabled, the tool reproduces the exact same game sequence on screen. Alternatively, disabling event mocking allows the user to interact with the game using normal keyboard input while preserving other aspects of the original execution context.

The tool supports comparative analysis across multiple recorded sessions. When a new replay is created, it is added to the timeline interface. The sliders from different sessions are temporally aligned from their starting indices, and when one game session is derived from another through replay, the timelines remain synchronized over their common execution period.

This synchronization enables developers to analyze differences between execution traces. Additionally, the interface includes a screen overlay feature that displays a semi-transparent (alpha 0.5) version of the screen from another execution, synchronized with the current timeline position. This visual comparison capability allows developers to immediately observe the effects of code changes on game behavior. The variables of the two executions are also displayed in the interface, and variable with different values are highlighted in green.

This tool supports spacetime exploration across time and code variants by enabling temporal navigation, replay from past frames under code changes with controlled external interactions, and synchronized comparison between original and replayed traces.

\subsubsection{Performance Evaluation}
\label{sec:flappy-performance}

We evaluated the performance overhead introduced by our monitoring system during Flappy Bird gameplay. Our measurements focus on two key metrics: frame rate impact and database storage requirements. As mentioned in Section~\ref{sec:storing-trace}, we utilize an in-memory database optimization during capture to minimize performance impact.

\begin{table}[t]
    \tiny
    \centering
    \begin{tabular}{|l|c|c|c|}
    \hline
    \textbf{Configuration} & \textbf{Time(FPS)} & \textbf{DB Size (Ko)} & \textbf{Overhead (\%)} \\
    \hline
    Unmonitored & 16.08s (60 FPS) & - & - \\
    Without screen capture & 16.07s(60 FPS) & 1.1 Mo & $\leq$1\% \\
    With screen capture as bitmap & 86.3s(11 FPS) & 5.7 Go & 530\% \\
    With screen capture as PNG& 16.08s(60 FPS) & 5.5 Mo & $\leq$1\% \\
    \hline
    \end{tabular}
    \caption{Performance impact of different monitoring configurations during Flappy Bird gameplay (1000 frames session)}
    \label{tab:flappybird_performance}
\end{table}

Table~\ref{tab:flappybird_performance} reports the performance of different tracing configurations on a recorded game session. The unmonitored configuration serves as the baseline. The three monitored configurations correspond to tracing without screen capture, tracing with direct capture of Pygame's internal frame representation as a bitmap, and tracing with the optimized PNG-based screen capture presented above. The results show that monitoring without screen capture, as well as monitoring with PNG capture, has no visible impact on the frame rate. This can be explained by the fact that the game speed is limited to 60 frames per second. In this setting, the monitored function is executed once per frame, and the additional tracing cost remains within the time available in the game loop, so it does not affect the observed frame rate. By contrast, storing frames in their raw bitmap representation introduces both a substantial runtime overhead and very high storage consumption. This result motivates the use of domain-specific trace representations and shows that practical usage may require tuning the captured data to the target application.

\subsubsection{Other Pygame development}

Although we only present Flappy Bird here, our repository includes other Pygame games (platformers and real-time strategy) that run with our tool with similar space-time support and required similar adaptation effort. 

\subsection{Omniscient Live Debugger in VSCode}
\label{sec:omniscient-live-debugger}

This subsection presents a VSCode extension for trace exploration integrated with the debugger.

The VSCode extension is activated on Python files. When one is opened, it searches the database for functions of the file that might have been monitored previously. If found, it adds a CodeLens (an interactive annotation in the code) next to the function. This CodeLens opens a view that allows users to see the different functions recorded, their call arguments, and return values.

If the recording has been done at line-level granularity, the extension allows exploration of the program's execution.

\begin{figure}[h]
    \centering
    \begin{subfigure}[t]{0.44\textwidth}
        \centering
        \includegraphics[width=\textwidth]{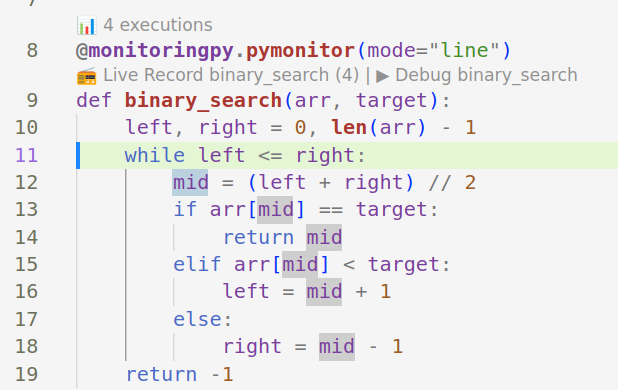}
        \caption{Code editor view showing the binary search function with highlighted execution line}
        \label{fig:vscode_code}
    \end{subfigure}
    \hfill
    \begin{subfigure}[t]{0.44\textwidth}
        \centering
        \includegraphics[width=0.9\textwidth]{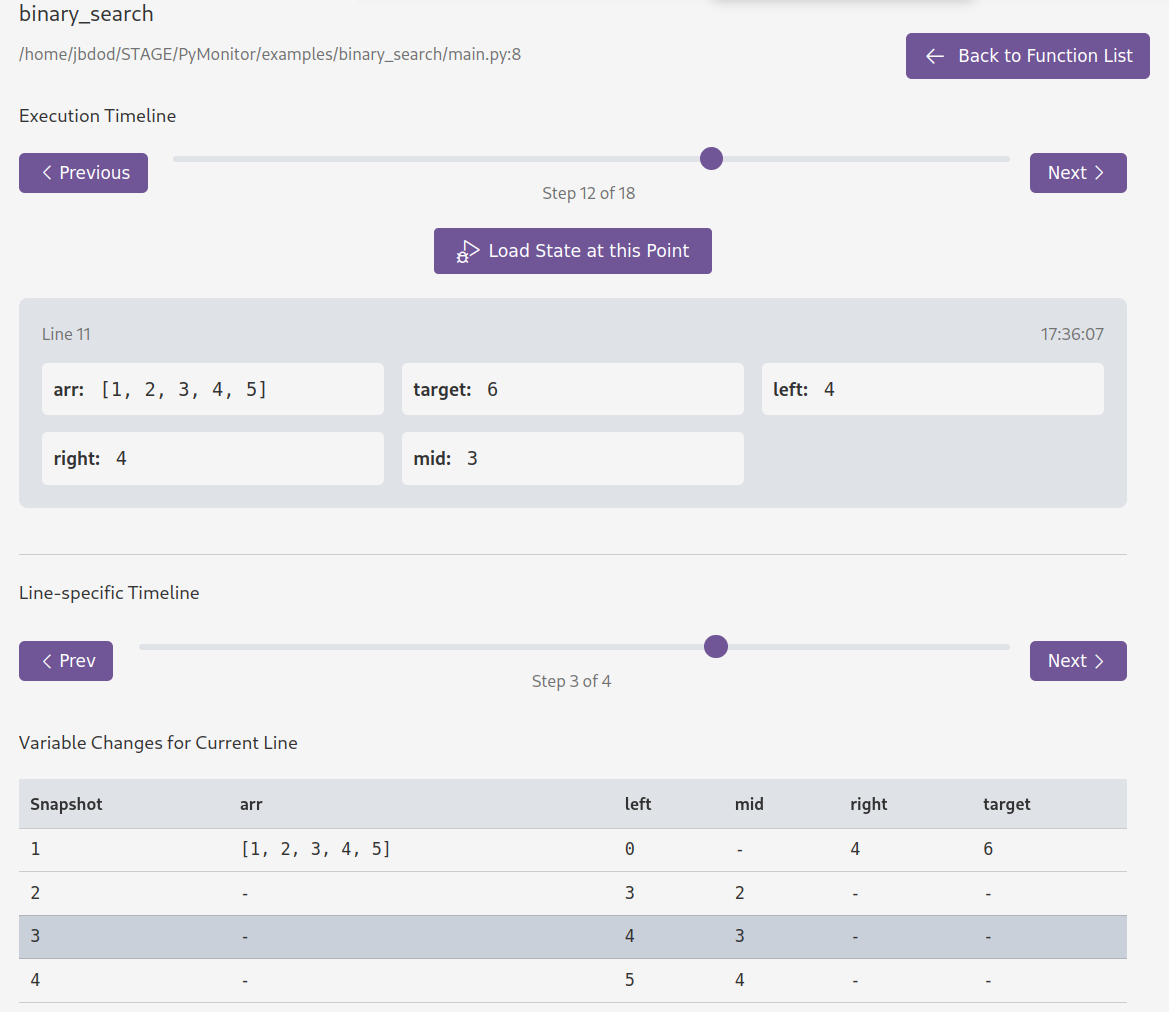}
        \caption{Exploration panel with execution timeline slider and line-specific timeline for variable tracking}
        \label{fig:vscode_exploration_panel}
    \end{subfigure}
    \caption{VSCode extension for execution exploration: highlighted code on the left and timeline/variable tracking panel on the right.}
    \Description{Two-panel VSCode screenshot: the left panel highlights the currently explored line in code, and the right panel shows timeline controls and variable values for execution exploration.}
    \label{fig:vscode_exploration}
\end{figure}

Figure~\ref{fig:vscode_exploration} shows a screenshot of VSCode while using the tool: here the function is inspected on a binary search function for the list \texttt{[1,2,3,4,5]} with target \texttt{6}. The left panel shows two parts: the execution timeline, which is a slider that allows exploration of the program's execution step by step, back and forward like in a debugger. The current line corresponding to the snapshot is highlighted in green in the editor, and the values are shown just below.

The second part is a line-specific timeline, which allows users to see the execution values related to the currently selected line and explore the steps recorded for a certain line in the code. This allows users to see, before the condition, how the values of the different variables (\texttt{left}, \texttt{mid}, \texttt{right}) change throughout the execution.

In addition to a view for exploring past execution, the extension allows interaction with traces while they are being recorded inside a debugger. The Debug CodeLens allows starting the debugger on a function, with input values from previous recorded executions (replay) or with default values.

\begin{figure}[t]
    \centering
    \includegraphics[width=0.85\textwidth]{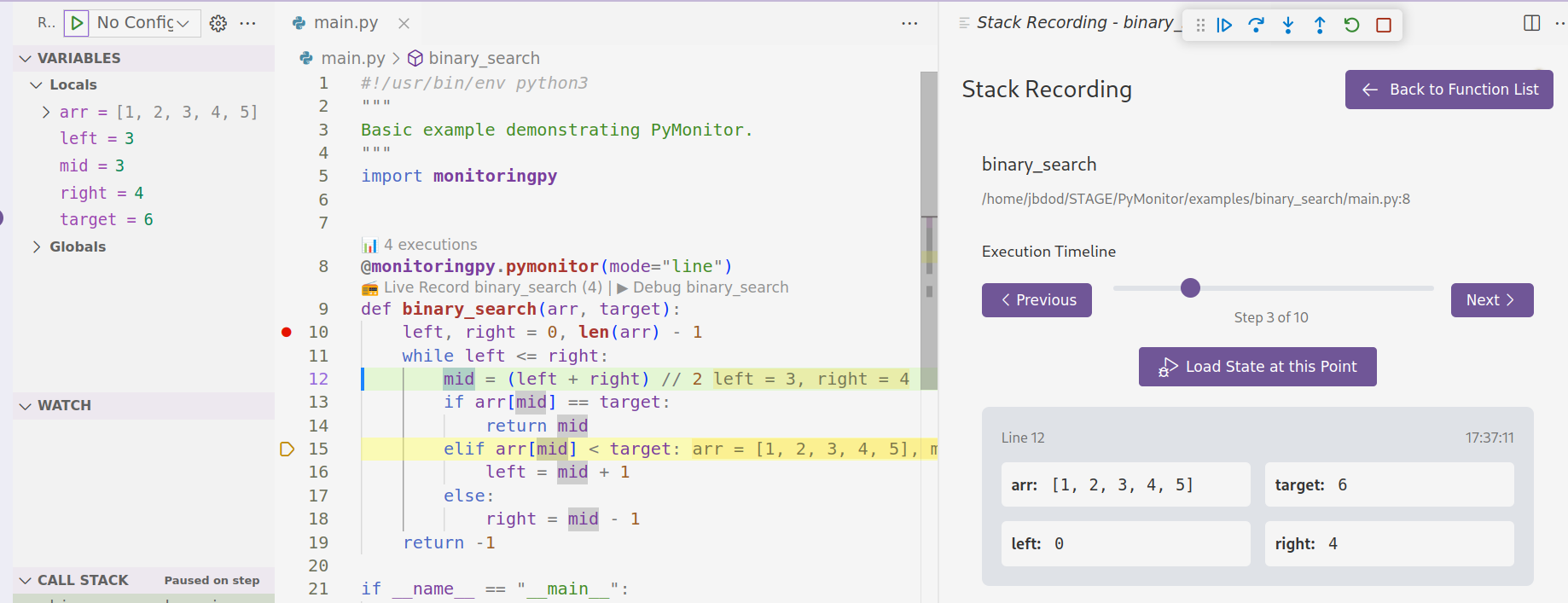}
    \caption{VSCode debugger integration with recorded-trace exploration: current debug state, trace focus, timeline slider, variable panel, and state-loading control.}
    \Description{VSCode debugger screenshot showing current execution position, recorded-trace focus, timeline slider, variable panel, and a control to load a past state into the running debug session.}
    \label{fig:vscode_debugger}
\end{figure}

Figure~\ref{fig:vscode_debugger} shows a screenshot of the VSCode screen while using the extension with the integrated debugger: the left part shows a yellow highlight in the code indicating the integrated debugger state. The debugger allows the use of breakpoints and step-by-step execution. The green highlight shows the line focus of the recorded execution, with slider and values on the right part. This allows users to see previous steps of the execution. In addition, the extension allows the possibility to return to a previous state by clicking on the \texttt{Load State at this point} button. The implementation loads the data recorded in the trace inside the Python interpreter and uses the Python debugger's capability to perform a goto jump to a specific line in the code.

The extension also provides code hot-swapping for Python at line-level granularity. It leverages trace information and a line mapping between the new version loaded in the Python VM and the previous one, computed using the GumTree\cite{DBLP:conf/icse/FalleriM24} diff algorithm, to inject updated code in the VM, and change code for active functions in the debugger. Combined with the state-loading functionality, this feature allows the developer to step back in time, load modified code, and experiment with different alternatives within the same debugging session. Finally, the exploration panel makes it possible to compare the values from a previous trace with those of another trace, or directly with the current execution. The alignment of states is also computed with the GumTree algorithm.

\begin{rqbox}{RQ1 \& RQ2}
Through the Pygame case study and the VSCode omniscient live debugger, we demonstrate that SpaceTimePy makes spacetime programming feasible (\textbf{RQ1}): it responds to the requirements of Section~\ref{sec:requirements} and tackles the technical challenges of Section~\ref{sec:spacetime-challenges}, supporting the trace cost vs.~fidelity trade-off, replay from recorded context under code substitution (state reinstantiation), and cross-variant alignment to compare executions over time. Both tools reuse the same trace/replay infrastructure and differ primarily in monitoring configuration and interface, demonstrating that the approach applies to different domains (\textbf{RQ2}).
\end{rqbox}

\subsection{Performance Evaluation}
\label{sec:performance}

This subsection reports tracing performance, first on HumanEval (164 problems) and then on five large Python projects from GitHub.

\subsubsection{Performance on HumanEval}

To evaluate the performance of SpaceTimePy on a larger set of examples, we applied it to the problems from the HumanEval dataset. The HumanEval dataset consists of 164 Python programming tasks, each with a reference solution and corresponding test cases. It is primarily used to benchmark the coding capabilities of large language models (LLMs). We selected this dataset because its problems are diverse and self-contained, representing typical functions one might wish to debug within a larger project. 

For each problem, we executed the test suite with a timeout of 180 seconds under three configurations: no monitoring, function-level monitoring, and line-level monitoring. We report in Table~\ref{tab:performance_humaneval} the number of problems that successfully passed the tests, the average execution time for each problem, and the size of the trace database.

\begin{table}[t]
\tiny
\begin{tabular}{lrll}
\toprule
Configuration & Problems passed & Avg. time per function(s) & Avg. DB Size (MB) \\
\midrule
Reference & 164 & 1.03e-03 & 0.00 \\
Function Monitoring & 164 & 3.03e-03 & 7.48 \\
Line Monitoring & 128 & 4.08e-02 & 8.69 \\
\bottomrule
\end{tabular}
\caption{Number of problems passed, execution time overhead (problem time), memory usage, and trace database size.}
\label{tab:performance_humaneval}
\end{table}

The results demonstrate that function-level monitoring introduces a 2-3x overhead on execution time, which is acceptable for development workflows where debugging and exploration are prioritized over raw performance. All 164 HumanEval problems pass with function-level monitoring.

Line-level monitoring introduces more significant overhead (approximately 40x slower execution), resulting in 36 failed test cases within the 180-second timeout. This demonstrates a practical trade-off: while line-level monitoring provides fine-grained execution visibility necessary for omniscient debugging, its performance cost limits applicability to computationally intensive functions. 

\subsubsection{Performance on Github popular Python projects}

\begin{table}[t]
\tiny
\centering

\setlength{\tabcolsep}{4pt} 
\renewcommand{\arraystretch}{1.2} 
\resizebox{\textwidth}{!}{%
\begin{tabular}{
l
p{0.8cm} 
p{0.9cm} p{0.5cm} 
p{0.9cm} p{0.5cm} 
p{1.1cm} 
p{1.3cm} 
p{1.1cm} 
p{1.0cm} 
p{1.6cm} 
}
\toprule
 & \multicolumn{1}{c}{\#Tests}
 & \multicolumn{2}{c}{Without SpaceTimePy}
 & \multicolumn{2}{c}{With SpaceTimePy}
 & \multicolumn{1}{c}{\#Serial.}
 & \multicolumn{1}{c}{Success (\%)}
 & \multicolumn{1}{c}{Time w/o (s)}
 & \multicolumn{1}{c}{Overhead (s)}
 & \multicolumn{1}{c}{DB Size (MB)} \\
\cmidrule(lr){3-4} \cmidrule(lr){5-6}
 & & Pass & Fail & Pass & Fail & & & & & \\
\midrule
beets & 1896 & 1896 & 0 & 1880 & 16 & 38759 & 91.0 & 31.14 & 28.49 & 16.73 \\
cherrypy & 289 & 281 & 8 & 278 & 11 & 36152 & 79.0 & 151.23 & 65.21 & 618.92 \\
discord.py & 282 & 282 & 0 & 282 & 0 & 16232 & 80.0 & 2.78 & 4.19 & 1.04 \\
dspy & 494 & 494 & 0 & 494 & 0 & 38364 & 88.0 & 60.48 & 22.78 & 29.18 \\
gensim & 996 & 996 & 0 & 993 & 3 & 442527 & 87.0 & 202.37 & 334.88 & 3671.09 \\
\bottomrule
\end{tabular}%
}
\caption{Performance of monitoring on real-world Python projects from GitHub.}
\label{tab:github_performance}
\end{table}

To evaluate the feasibility of SpaceTimePy on real-world Python projects, we sampled popular and large open-source repositories from GitHub. The selection criteria required that Python be the primary language, with each project having over 1,000 stars, more than 2,000 commits, and at least 200 contributors. From this pool, we randomly selected five repositories with a test suite runnable with \texttt{pytest}.

We compared the execution of the test suites without monitoring and with line-level monitoring of test cases. This setup was chosen to simulate a realistic development scenario in which test cases serve as examples for debugging sessions.

The selected projects cover a broad range of application domains, including web development, bot scripting, topic modeling, music management, and AI frameworks.

Table~\ref{tab:github_performance} presents the results of this experiment. For each repository, we report the number of tests passed with and without monitoring, the execution time of the test suite without monitoring, the additional execution time introduced by monitoring, the size of the trace database, and the success rate of variable serialization.
As explained in Section~\ref{sec:storing-trace}, non-serializable variables are excluded from the trace when no serialization method is available.

The results reveal several important insights about the practical applicability of our approach. For most projects, the majority of tests pass successfully with monitoring enabled, indicating that the trace system is usable in typical development scenarios.

Approximately 80\% of objects are serializable without requiring custom serialization code, demonstrating that the default serialization mechanisms handle most common data structures effectively. However, this also indicates that approximately 20\% of objects require custom handling, which aligns with the serialization challenges discussed in Section~\ref{sec:serialization-challenges}.

The execution time overhead varies significantly across projects, ranging from approximately 35\% to 150\% depending on the project characteristics. This variation reflects differences in the computational intensity of test suites and the amount of state captured during execution.

Memory usage exhibits even greater project-dependent variation, with trace database sizes ranging from approximately 1 MB to several gigabytes. The substantially larger memory footprint observed in the gensim project can be attributed to the nature of its workload: natural language processing and topic modeling tasks typically involve processing and storing large volumes of textual data, which significantly increases the volume of state that must be captured and serialized.

These results demonstrate that the performance impact of SpaceTime Programming is highly dependent on the characteristics of the target project.

\begin{rqbox}{RQ3}
Function-level monitoring introduces a 2--3$\times$ overhead and is compatible with all 164 HumanEval problems, making it practical for interactive development workflows. Line-level monitoring incurs a higher cost ($\approx$40$\times$), which can cause timeouts on computation-intensive functions but remains usable for typical debugging targets. On real-world GitHub projects, the overhead ranges from 35\% to 150\% in execution time, with approximately 80\% of objects serializable out of the box. Memory usage varies widely depending on the volume of captured state. Overall, the tracing overhead is acceptable for interactive use at function-level granularity and viable at line-level granularity when applied selectively to functions of interest.
\end{rqbox}

\section{Discussion}
\label{sec:discussion}

This section discusses implications of SpaceTime Programming in terms of implementation trade-offs, applicability, and cognitive load.

\subsection{Implementation Trade-offs}
\label{sec:serialization-challenges}
The instrumentation required for SpaceTime Programming introduces a runtime overhead that can affect the usability of the resulting tools. This directly corresponds to the \emph{trace cost vs. fidelity} challenge from Section~\ref{sec:spacetime-challenges}. As our performance evaluation highlights, overhead depends heavily on instrumentation granularity and the nature of the captured data. While function-level monitoring generally remains within acceptable bounds for interactive development, finer granularities, such as line-level tracing, can incur significant costs in computation-intensive scenarios.

Mitigating this overhead requires careful optimization. Adjusting the instrumentation granularity allows developers to balance the level of detail with execution performance. Furthermore, serialization presents both challenges and opportunities for optimization. Default serialization mechanisms often fail to handle complex objects, such as closures, necessitating the implementation of custom serialization logic or manual intervention to exclude specific objects. 

However, domain-specific optimizations can yield substantial improvements. For instance, in our Pygame case study, we utilized a compressed image format for screen captures (instead of the original bitmap representation), which drastically reduced both the database size and the serialization overhead. Other strategies, such as trace compression or on-demand state regeneration (recomputing states during exploration rather than capturing them exhaustively), represent promising avenues for further reducing the footprint of the system.

\subsection{Applicability Considerations}

We demonstrate \spacetime through SpaceTimePy, but the approach is not specific to Python. More generally, implementing the trace model and operations of Section~\ref{sec:trace-definition} and Section~\ref{sec:trace-operations} requires the capabilities identified in Section~\ref{sec:required-pl-ft}. Any language/runtime that provides these capabilities, via instrumentation, debugging interfaces, or runtime support, is a candidate for a \spacetime system.

In practice, these capabilities may be only partially available or may not be available uniformly at all granularities. For example, a runtime may expose reliable function-level call/return monitoring but not efficient statement-level stepping, or it may allow observing values but not reinstantiating execution from an arbitrary point. Such granularity-dependent limitations directly constrain what forms of replay, code substitution, and cross-variant comparison are feasible.

While our prototypes target an imperative setting, other paradigms should also be compatible in principle as long as executions can be represented as traces. In a functional programming language, higher-order functions and closure capture could complicate both observation and serialization, and may require adapted representations or runtime support.

Finally, applicability depends not only on runtime support but also on tool-building effort: tool builders must define capture points and state migration strategies, provide controls for managing variants, and align executions across time. These concerns mirror the \emph{variant management and comparison} and \emph{cross-variant alignment} challenges from Section~\ref{sec:spacetime-challenges}, and remain key determinants of end-to-end usability even when tracing and replay are technically feasible.

\subsection{Cognitive Load}

Simultaneous space-time exploration also raises usability concerns. While finer control over execution and code variants broadens exploration, it also increases cognitive load. This is the human-facing counterpart of the technical challenges above: as the system makes branching, replay, and comparison easier, it also increases the number of artifacts (states, variants, alignments) that developers must interpret. The literature on human-centric computing suggests that overloading developers with unfiltered information can be counter-productive \cite{6227188}.

Although this work focuses on the technical feasibility and applicability of the SpaceTime Programming, the presentation of information is critical to its adoption. It is essential to design interfaces that present complex execution data without overwhelming the user. Future research must address these usability challenges by investigating visualization techniques that effectively manage this complexity and by validating the utility of the approach through user studies.

\section{Related work}
\label{sec:related-work}

Omniscient debugging has been extensively explored as a means to enable backward navigation through program execution history. Bousse et al.~\cite{bousse_omniscient_2018} present comprehensive techniques for constructing execution traces and exploring the evolution of values across trace histories, while Leroy et al.~\cite{leroy_trace_2018} propose operations for comparing and merging execution traces. These approaches excel at temporal exploration but operate under the assumption of static code during trace capture and analysis. Tsai~\cite{tsai_noninterference_1990} presents methods for capturing program state to enable replay for bug reproduction, focusing on low-level processor mechanisms for real-time systems, but does not address replay scenarios where code has been modified between the original execution and the replay attempt. Our SpaceTime Programming approach extends this foundation by explicitly incorporating code versioning into the trace system, enabling developers to understand not only how programs evolve over time but also how code changes affect execution behavior.

Live programming environments aim to provide immediate feedback on code changes through continuous execution and state visualization. Recent work by Döderlein et al.~\cite{doderlein_liverec_2024} presents LiveRec, a tool for exploring program state through probes, though it focuses primarily on stack variables and lacks support for global state exploration. Their approach is limited by reliance on Debug Adapter Protocol (DAP) accessibility, which constrains both performance and the scope of accessible program state. In contrast, our SpaceTimePy implementation leverages Python's \texttt{sys.monitoring} capabilities to achieve significantly better performance while providing comprehensive access to both local and global program state.
Work on integrating runtime data into development environments includes Cito et al.~\cite{cito_runtime_2015}, who propose mechanisms for recording and displaying variables in live development contexts, and Winter et al.~\cite{winter_monitoring-aware_2019} and Chatley et al.~\cite{chatley_supporting_2019}, who extend this concept to monitoring-aware IDEs that support development experiences with production metrics. While these approaches excel at capturing and visualizing runtime information to enhance development awareness, they primarily focus on monitoring for analysis rather than leveraging captured data for future execution scenarios or code exploration.

Exploratory programming and time-travel debugging represent complementary approaches to program exploration. Beckmann et al.~\cite{beckmann_probing_2025} present an approach for exploring multiple program variants simultaneously within the Smalltalk/Squeak live programming environment, enabling comparison of different program versions but lacking the temporal exploration capabilities that allow developers to understand how variants behave over time. Salkeld~\cite{salkeld_time-travel_2018} proposes an omniscient debugger that allows developers to add and invoke new code within the context of historical traces, enabling interaction with past program states. This approach focuses on executing new code that can interact with previously captured objects rather than replaying execution with modified code.

Recent work by Tiwari et al.~\cite{tiwari_production_2022,tiwari_proze_2024,tiwari_mimicking_2024} explores techniques for mocking functions based on previous execution data to automate test generation. This approach focuses on capturing function behaviors from production executions and replaying them in test environments to improve test coverage and reliability. Like our approach, they leverage function mocking based on previous execution data, though they target automated testing scenarios rather than interactive development exploration.

\section{Conclusion and future work}
\label{sec:conclusion}

This paper introduced SpaceTime Programming, a unified approach that bridges the gap between live programming, exploratory programming, and omniscient debugging by enabling simultaneous exploration of both code variants (space) and execution flow (time). Our key contribution is the design of this new model and the demonstration of its technical feasibility.

We implemented SpaceTimePy using \texttt{sys.monitoring} and demonstrated it through two case studies: a Pygame development environment and a live omniscient debugger. These prototypes show how developers can capture sessions, explore execution timelines, and iteratively refine code while preserving execution context.
We further evaluated it on the HumanEval benchmark and on five large real-world Python projects from GitHub. Results showed that on real-world projects, the execution time overhead ranged from 35\% to 150\%, with approximately 80\% of objects being serializable out of the box.

This work opens multiple avenues for future research. Technical improvements, such as optimized serialization and reduced monitoring overhead, would further enhance applicability. The approach could also be extended to other programming languages and Domain-Specific Languages (DSLs) to demonstrate generalizability, and to other programming paradigms (e.g., functional languages) to validate the model beyond the imperative setting studied here. Finally, future studies should focus on user experience to evaluate how developers interact with these tools and to validate the utility of simultaneous space and time exploration in practice.

\section{Data-Availability Statement}
The implementation of SpaceTimePy, including the Pygame exploration tool, is available at \url{https://anonymous.4open.science/r/SpaceTimePy-BC0F/}.
The VSCode extension is available at \url{https://anonymous.4open.science/r/spacetimepy-vscode-2B48/}.
The code and data for the performance evaluation on GitHub projects are available at \url{https://zenodo.org/records/19062251}. The supplementary material includes a video demonstration of the Pygame exploration tool.

\bibliographystyle{unsrtnat}
\bibliography{sample}

\end{document}